\documentclass[mathleft]{an}
\usepackage{graphicx}
\usepackage{times}
\usepackage{delarray}
\Pagespan{313}{}
\Yearpublication{2007}%
\Month{3}%
\Volume{328}%
\Issue{3-4}%
\DOI{10.1002/asna.200610736}%

\def\bxi{\mbox{\boldmath$\xi$}}

\def\bU{\mbox{\boldmath$U$}}

\def\br{\mbox{\boldmath$r$}}

\def\id{{\rm d}}
\def\br{\mbox{\boldmath$r$}}

\def\unitx{\mbox{\boldmath$\hat{x}$}}
\def\unitz{\mbox{\boldmath$\hat{z}$}}

\def\bu{\mbox{\boldmath$U$}}
\def\bnabla{\mbox{\boldmath$\nabla$}}
\def\bJ{\mbox{\boldmath$J$}}
\def\bB{\mbox{\boldmath$B$}}
\def\bnabla{\mbox{\boldmath$\nabla$}}

\newcolumntype{L}{>{$}l<{$}}
\newenvironment{Cases}{\begin{array}\{{lL}.}{\end{array}}

\begin{document}

\sloppy

\title{SLiM: a code for the simulation of wave propagation\\
through an inhomogeneous, magnetised solar atmosphere}
\author{R. Cameron\thanks{\email{cameron@mps.mpg.de}} 
\and L. Gizon \and K. Daiffallah}
\titlerunning{Simulation of wave propagation
through an inhomogeneous, magnetised solar atmosphere}
\authorrunning{Cameron et al.}
\institute{Max-Planck-Institut f\"ur Sonnensystemforschung, 
37191 Katlenburg-Lindau, Germany}

\received{24 Nov 2006}
\accepted{}
\publonline{later}


\abstract{
In this paper we describe the semi-spectral linear MHD (SLiM)
code which we have written to follow the interaction of linear waves 
through an inhomogeneous three-dimensional solar atmosphere. 
The background model allows almost arbitrary perturbations of density, temperature, sound speed as well as magnetic and velocity fields. We give details of several of 
the tests we have used to check the code. The code will be useful in 
understanding the helioseismic signatures of various solar features, including sunspots.}  
\maketitle

\section{Introduction}
Local helioseismology uses waves propagating through the solar interior 
to probe the inhomogeneities therein. Flows, temperature  
and magnetic field inhomogeneities all imprint their signatures on 
the waves, and helioseismology attempts to learn about the 
inhomogeneities by studying the waves (Duvall et al. 1993). In order to better 
understand the various signatures we have developed a numerical code 
for forward calculations which numerically propagates waves through 
various types of inhomogeneities.

Such calculations have been performed in the past, but mainly in a
2-D geometry (eg Cally \& Bogdan 1997), or with other simplifying physical
assumptions (eg Bogdan et al. 1996; Fan, Braun \& Chou 1995).
Analytic solutions also exist for some idealised
problems: sound speed perturbations (eg Jensen, \& Pijpers 2003),
an isolated mass flow (Birch \& Felder 2004), or a flux tube in an 
otherwise uniform atmosphere (Gizon, Hanasoge, \& Birch 2006). 
These solutions have been very useful
in revealing various effects and in aiding our understanding. 
There remains however a need for 3-dimensional
treatments (see Werne, Birch \& Julien 2004).

This paper describes the three-dimensional physical and mathematical 
problem we have chosen to tackle, as well as the numerical scheme we have 
implemented. The code has been subjected to a number of tests, several
of which will be described in detail. Our intention is to make the code 
available to the helioseismology community as part of the HELAS 
project. 

\section{Statement of the problem}

\subsection{The nonlinear equations}
As our starting point we consider a typical formulation of the MHD equations
in cgs units. The relevant equations are the continuity equation,
\begin{eqnarray}
\partial_t \rho  + \bnabla \cdot (\rho\bu)=0\,, 
\end{eqnarray}
the compressible Navier-Stokes equation including Lorentz and buoyancy forces,
\begin{eqnarray}
\rho D_t \bu  &=&
          -\nabla P + \rho g \unitz
          + \frac{1}{4 \pi} (\bnabla \times \bB) \times \bB \nonumber \\
          && + \mu \Delta \bu + \frac{\mu}{3} \bnabla ( \bnabla\cdot\bu)   \,,
\end{eqnarray}
where $D_t = \partial_t +\bu\cdot\bnabla$, the induction equation,
\begin{eqnarray}
\partial_t\bB =\bnabla \times (\bu \times \bB)+\eta
\Delta \bB\,,
\end{eqnarray}
an energy equation,
\begin{eqnarray}
\rho C_v D_t T
             &=&-P\bnabla \cdot\bu +K \nabla ^2 T  \nonumber \\
&& +\mbox{viscous heating}+\mbox{ohmic heating} \nonumber \\
&&+\mbox{radiative heating} 
\, ,
\end{eqnarray} 
an equation of state, $P=P(\rho,T)$, and the solenoidality condition
$\bnabla \cdot\bB=0$. Usual symbols are used to denote the physical quantities that appear in the above equations. The gravitational acceleration, $g$, is assumed to be uniform and constant. The diffusivities as written here are particularly simple, being both
isotropic and  uniform, and in principle more complicated forms might
be needed in certain applications, however as we intend to treat the waves
as ideal perturbations and ignore the effects of the diffusivities, the
particular forms of these terms is not important.

\subsection{The ideal, linear equations}

The waves are generally small perturbations, excited by the turbulent
motions of the granules or in rare cases as a result of impulsive events 
such as flares. Our code follows the evolution of a such perturbation 
superimposed on a provided background. The background, assumed to be 
time-independent and in 
equilibrium, and is specified by  giving the density $\rho_0(\br)$, 
pressure $P_0(\br)$, 
magnetic field ${\bB_0}(\br)$, velocity field
${\bU_0}(\br)$ and the first adiabatic index $\Gamma_1(\br)$ at position $\br=(x,y,z)$  everywhere inside the domain. 
The background current, $\bJ_0 = \bnabla \times \bB_0$, and sound speed, $c_0 = (\Gamma_1 P_0 /\rho_0)^{1/2}$, can be derived from the above quantities.

The perturbation is assumed to be sufficiently small that we can consider 
the linearised equations. 
We currently assume the waves are adiabatic and the 
wave propagation is ideal. This allows us to write the magnetic, velocity, 
pressure and density perturbations in terms of the displacement vector $\bxi$. Using primed quantities as the Eulerian perturbations to the background 
state, the master equation for $\bxi$ is 
\begin{eqnarray}
\rho_0 \partial_t^2 \bxi &=&-\nabla  P'+ \rho' g \unitz \nonumber 
+ \frac{1}{4\pi} (\bJ'\times \bB_0 + \bJ_0 \times \bB')
                                              \nonumber \\
& &+ (\rho_0 \partial_t \bxi-  \rho' \bU_0 - \rho_0 \bU')\cdot \bnabla  \bU_0 \nonumber \\
& &- \rho_0 \bU_0 \cdot \bnabla (\partial_t\bxi+\bU') ,
\end{eqnarray}
where the perturbed quantities that appear in this equation are themselves functions of $\bxi$ according to
\begin{eqnarray}
\rho'&=&-\bnabla \cdot (\rho_0 \bxi) \label{eq.con} ,\\
P'  &=& c^2_0  ( \rho' + \bxi \cdot \bnabla \rho_0 ) 
        -\bxi \cdot \nabla P_0  ,\\
\bB'&=&\bnabla \times (\bxi \times \bB_0) \label{eq.ind},\\
\bJ'&=&\bnabla \times \bB' ,\\
\bU'&=&\partial_t \bxi+ \bU_0 \cdot \bnabla\bxi 
                          - \bxi \cdot \bnabla\bU_0 \label{eq.disp} .
\end{eqnarray}
We refer the reader to Lynden-Bell and Ostriker (1967) for equation~(\ref{eq.disp}), to Christensen-Dalsgaard (2003, page 174) for a treatment of the continuity equation (eq.~[\ref{eq.con}]), and to Moffatt (1978) for the induction equation (eq.~[\ref{eq.ind}]). Note that we have assumed that the presence of a flow and a magnetic field are mutually exclusive (eq. [\ref{eq.ind}]).

\subsection{The boundary conditions}

The tests described in this paper involve magnetic, density or velocity 
inhomogeneities superimposed on an isotropic, uniform background. For
these tests we have then assumed periodicity in all three spatial
directions. However since the code is also intended to be used for
stratified atmospheres, we will here describe the boundary condition we
have implemented for such cases.

We only intend to simulate small regions of the solar
surface and so have used a box-geometry and used a spectral treatment
in the horizontal directions and finite difference for the vertical.
The boundary conditions are naturally important. The side boundaries are
simple: the box is periodic. For the upper boundary we currently use
the condition that the Lagrangian perturbation of the vertical component
of the stress tensor vanishes (following Cally and Bogdan 1997):
\begin{eqnarray}
\Pi'_{iz} +  \bxi \cdot \bnabla \Pi_{iz} =0 ,
\end{eqnarray}
where $\Pi$ is the stress tensor.
Furthermore we assume that the background flow is restricted to regions
away from the top boundary, and that there is no purely horizontal field 
there, so that  
\begin{eqnarray}
\Pi_{ij}=[P+B^2/(8\pi)]\delta_{ij}-B_i B_j/(4 \pi).
\end{eqnarray}
In regions where $\bB_{0} \ne {\bf 0}$ the condition
$\Pi'_{xz} +  \bxi \cdot \bnabla \Pi_{xz} =0$
implies
\begin{eqnarray}
B_x'&=&-\frac{B_{0x} B_z'+ \bxi \cdot \bnabla(B_{0x}B_{0z})}{B_{0z}}
\end{eqnarray}
and the condition on $\Pi'_{yz}$ implies 
\begin{eqnarray}
B_y'&=&-\frac{B_{0y} B_z'+ \bxi \cdot \bnabla (B_{0y}B_{0z})}{B_{0z}} .
\end{eqnarray}
The boundary condition
$\Pi'_{zz} +  \bxi \cdot \bnabla \Pi_{zz} =0$
can then be converted into a constraint on $P'$ at the top boundary:
\begin{eqnarray}
P'&=& \frac{1}{4\pi}(B_{0z} B_{z}' - B_{0x}B_{x}'-B_{0y} B_{y}')
- \bxi \cdot \bnabla P_0 \nonumber\\ 
  & & 
          -  \frac{1}{8 \pi} \bxi \cdot \bnabla (B_{0x}^2+B_{0y}^2-B_{0z}^2) .
\end{eqnarray}
Once this is known we can also deduce the density at the top boundary using 
equation~(7):
\begin{eqnarray}
\rho'&=& c_0^{-2} (P'+\bxi \cdot \bnabla P_0) 
        - \bxi \cdot \bnabla \rho_0 .
\end{eqnarray}
Lastly, our assumption that the velocity vanishes near the boundary implies
\begin{eqnarray}
\bU'&=&\partial_t \bxi .
\end{eqnarray}
In those regions where $\bB_0={\bf 0}$, the boundary conditions have $\bB'=0$ but
are otherwise unchanged.

In the case where there is a magnetic field at the boundary, 
it can readily be seen from equation~(8) that specifying $B'_{x}$ on the 
boundary is equivalent to specifying a relationship between 
$\partial_z \xi_x$ and $\partial_z \xi_z$.  
Similarly a constraint on 
$\partial_z \xi_y$ and $\partial_z \xi_z$ is given by the condition on 
$B'_{y}$ and all three are related by $\partial_z \xi_z$ by the condition on 
$\rho'$ and equation~(6). We thus have three relationships for three unknowns
and have a well posed problem for defining ${\partial_z}{\bxi}$:
the boundary conditions, in magnetic regions, are equivalent to
specifying a boundary condition three components of ${\partial_z}{\bxi}$.

In the non-magnetic there are fewer wave modes and the boundary condition
reduces to specifying ${\partial_z}{\xi_z}$. In this special case the
boundary condition corresponds to that of a free-surface 
(Cally and Bogdan 1997).

\subsection{Initial Conditions}
The above equations allow us to follow the evolution of arbitrary 
perturbations $\bxi(\br,t_0)$ and $\partial_t\bxi(\br,t_0)$. Often these 
initial conditions will take the form of an f- or p-mode packet, but our
code is able to deal with arbitrary initial perturbations.

\subsection{Physical instabilities}
The code evolves perturbations superimposed on the background atmosphere
assuming that the evolution is linear. Consequently if the atmosphere
is (linearly) unstable then exponentially growing modes exist. Under almost
all circumstances the fastest growing of these modes will 
eventually dominate the solution. This is particularly important to 
remember when our interest is wave propagation in the convection zone
where the atmosphere is in places superadiabatically stratified. 
In such places the atmosphere is convectively unstable, and exponentially
growing solutions are to be expected: this is a real property
of the background, and will necessarily exist in any proper numerical
treatment of the problem.

Our suggestion, and the approach we have adopted in some
preliminary calculations, is to preprocess the background so that it becomes 
marginally stable. This removes the unstable modes without introducing new 
unphysical modes, for example gravity waves in the convection zone.
Naturally the code can also be used to study the linear growth of ideal
instabilities without modification to either the code or the 
atmosphere.

\section{Numerical Methods}
Since we have chosen periodic boundary conditions for the side walls,
we have implemented a pseudo-spectral scheme in the horizontal directions. 
That is all horizontal derivatives are evaluated in Fourier space, all 
products are evaluated in physical space: we use the FFTW package (Frigo \&
Johnson 2005) to move between these two spaces.
Since our aim is to model the 
Sun, stratification due to gravity implies periodic boundaries are not
appropriate for the top and bottom boundaries. We have therefore used
a simple two-step Lax-Wendroff scheme in the vertical to evolve the 
horizontal Fourier modes. Care has been taken in order to ensure
that the various derivatives are defined on the appropriate grids.

Since the Lax-Wendroff scheme is reasonably efficient in damping waves 
with wavelengths comparable to the grid-spacing, we have implemented
a hyper-diffusivity in the horizontal directions which scales  like $k^4$,
where $k$ is the horizontal
wavenumber. The short wavelength modes in the horizontal directions 
are thus damped in a similar way to that which occurs in the
vertical without significantly affecting those
modes we are interested in.

\section{Comparison with exact solutions}
Here we provide some examples of test calculations used to verify the code.
For these tests, the background is uniform away from the inhomogeneity and the initial condition is a simple acoustic wave packet propagating in the $+\unitx$ direction (to the right): 
\begin{equation} 
\bxi (\br, t_0) = - \unitx \int_0^{\infty} A(k) \, \sin[k(x-x_0)] \,  \id k  ,
\end{equation}
and
\begin{equation}
\partial_t \bxi (\br, t_0) = \unitx \int_0^{\infty}   A(k) \, c_0 k \, \cos[k(x-x_0)] \, \id k  ,
\end{equation}
where $\br=(x,y,z)$ and $A(k)$ is a Gaussian envelope centred on the wavenumber corresponding to angular frequency $3$~mHz and with standard deviation $1$~mHz. At $t=t_0$ the wave packet is centred at the left edge of the computational domain with $x_0=-12$~Mm.

\subsection{Magnetic and pressure perturbations}

\begin{figure}
\resizebox{0.95\hsize}{!}{\rotatebox{0}
             {\includegraphics{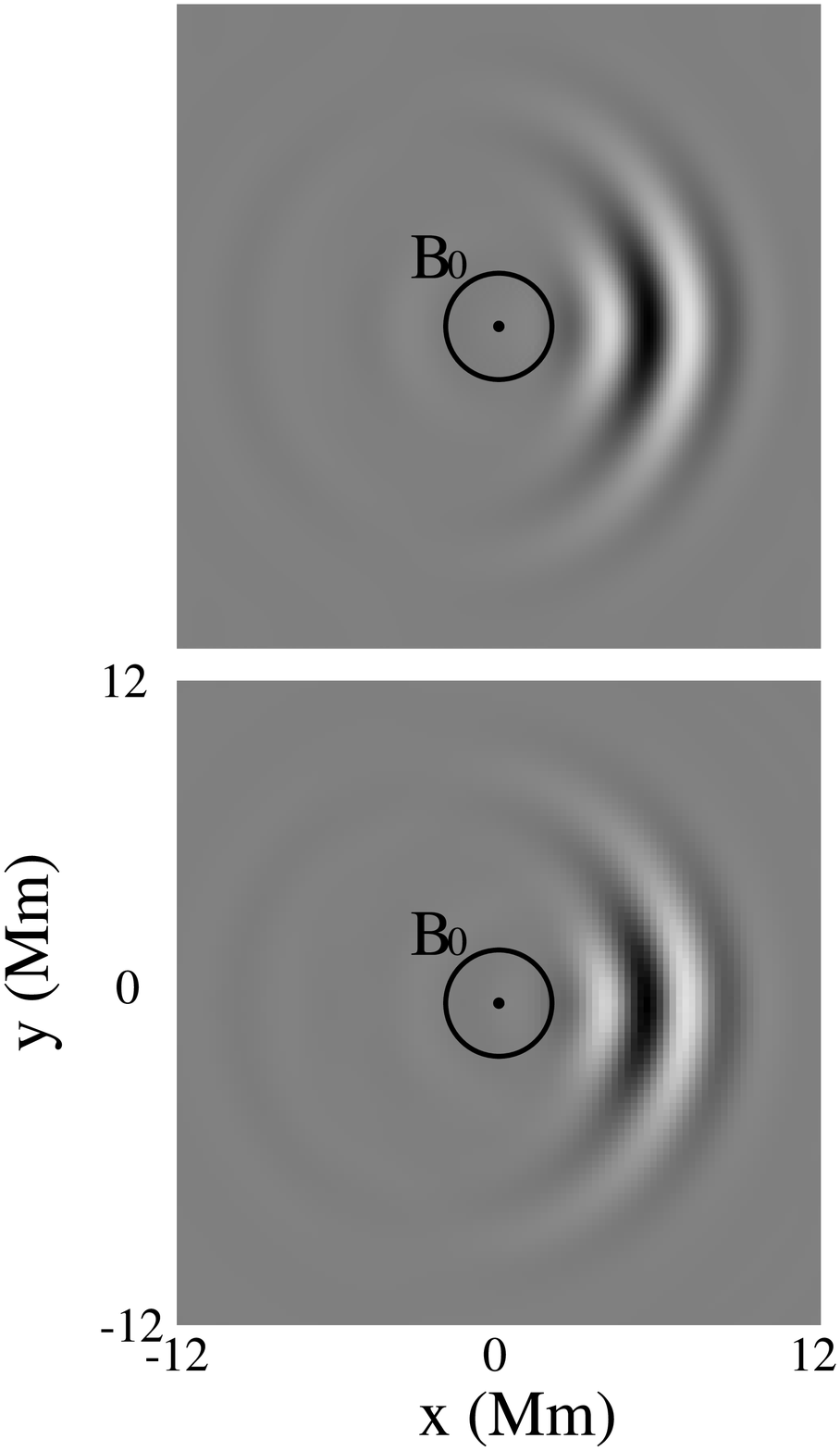}}}
\caption{Scattered pressure perturbations (snapshot) from a vertical 1~kG flux tube with axis $x=y=0$. The radius of the tube is 2~Mm. 
The bottom panel shows the exact result and the top panel
 the numerical result. }
\label{fig:h1kG}
\end{figure}

\begin{figure}
\resizebox{0.95\hsize}{!}{\rotatebox{-90}
             {\includegraphics{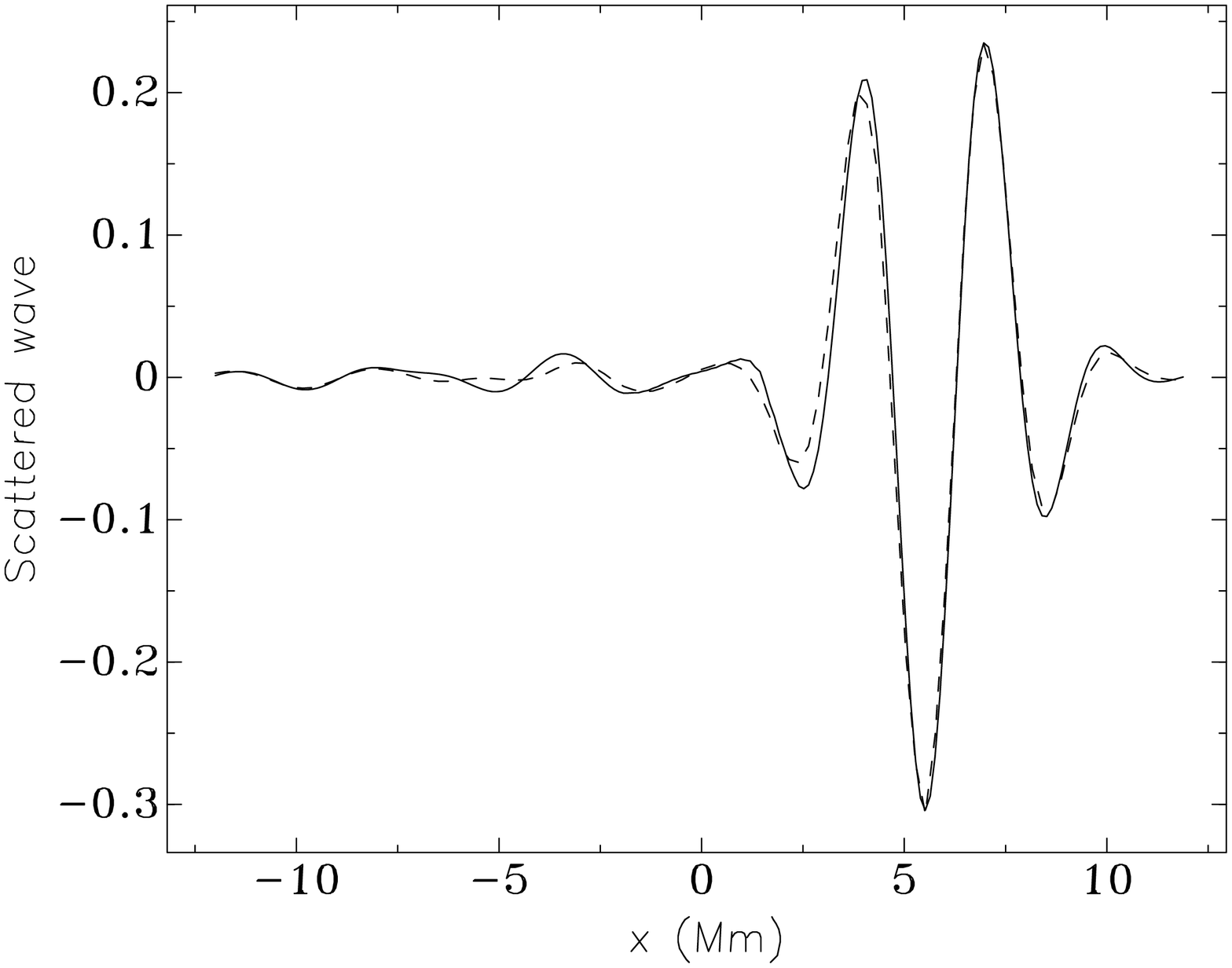}}}
\caption{A cut at $y=0$ through the scattered pressure perturbations shown in Figure~\ref{fig:h1kG} (vertical 1~kG flux tube). The dashed line shows the exact solution and the solid line the numerical. The units are such that the incoming pressure perturbations have an amplitude of 1.}
\label{fig:h1kGslice}
\end{figure}

\begin{figure}
\resizebox{0.95\hsize}{!}{\rotatebox{0}
             {\includegraphics{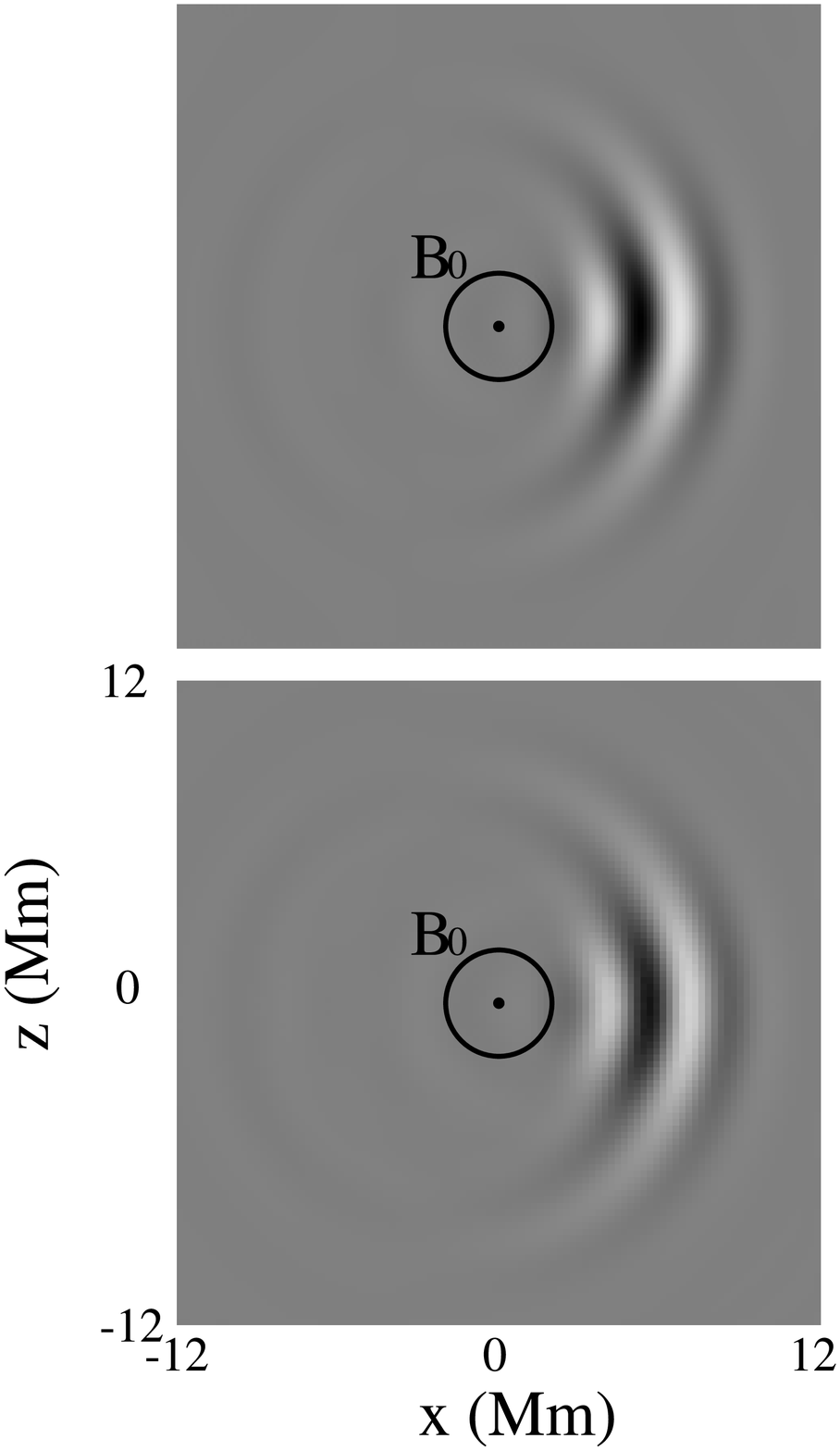}}}
\caption{
Scattered pressure perturbations (snapshot) from a horizontal 1~kG flux tube with axis $x=z=0$. The radius of the tube is again 2~Mm. 
The bottom panel shows the exact result and the top panel
 the numerical result.}
\label{fig:v1kG}
\end{figure}

Our first test reproduces the analytic result of Wilson (1980) and Gizon et al. (2006) where the problem is fully described. 
In brief, the background state consists of a cylindrical 
flux tube in pressure and thermal equilibrium. The external plasma is  
to be uniform and at rest, with density $5\times 10^{-7}$~g/cm$^3$, 
sound speed  $11 \times 10^5$~cm/s, and with $\Gamma_1=5/3$. The flux tube,
which has a radius of 2~Mm in our tests, has uniform properties. Its magnetic
field is uniform, and we have considered different field strengths 
$B_\mathrm{tube}$ from 1~kG
(relatively weak) to 3~kG (where the flux tube is 99\% evacuated). The magnetic
field and the condition of pressure equilibrium imply that the pressure
inside the tube is
\begin{equation}
P_{\mathrm{in}}=P_{\mathrm{ext}}- B_{\mathrm{tube}}^2/(8 \pi). 
\end{equation}
The assumed thermal equilibrium requires that the temperature inside the 
tube equals that outside the tube and that the density inside the tube
be reduced.

The problem thus has a discontinuous jump across the flux tube. This jump
is important for obtaining the analytic solution but is somewhat problematic
for a numerical treatment. Fortunately there is no reason for expecting
the discontinuity to be singular, and we can therefore proceed by
considering the discontinuity as the limit of a thin boundary layer
over which the properties of the tube vary. Specifically we assume
\begin{equation}
B_0 = B_{\mathrm{tube}} \begin{Cases}
    1 & if  $r \le r_0$,\\
    (r_1-r)/(r_1-r_0) & if $r_0<r<r_1$, \\
    0 & if $r_1 \le r$ ,\\
     \end{Cases}     
\end{equation}
where $r$ is the distance from the tube axis,
and consider the limit $r_1 \rightarrow r_0$. In practice we
resolve $r_1-r_0$ using 5 grid points and increase the resolution until
the behaviour converges. In any case we choose $r_0$ and $r_1$ such that 
the total flux corresponds to a 2~Mm tube with field strength 
$B_{\mathrm{tube}}$.

This completes the specification of the background. On this background
we consider a plane parallel wave. The test cases considered here assume that
the direction of propagation is perpendicular to the flux tube so that
the problem is essentially two-dimensional. Since we use different schemes 
in the horizontal and vertical directions we have checked the code using 
vertical and horizontal tubes. Figures~\ref{fig:h1kG} and ~\ref{fig:h1kGslice}
show a comparison of the scattered pressure perturbations computed analytically and numerically for a vertical 1~kG flux tube. The numerical
calculation used $200$ Fourier modes in both the $x$ and $y$ directions.
Similar results were found for the horizontal tube (see Fig~\ref{fig:v1kG}).

It is important to comment that the resolution (24~Mm over 200 pixels 
in each direction) is required to accurately capture the wave interaction
with the discontinuities at the edge of the tube rather than the
tube itself. For problems without discontinuities we expect convergence
will occur at even lower resolutions.

\begin{figure}
\resizebox{0.95\hsize}{!}{\rotatebox{0}
             {\includegraphics{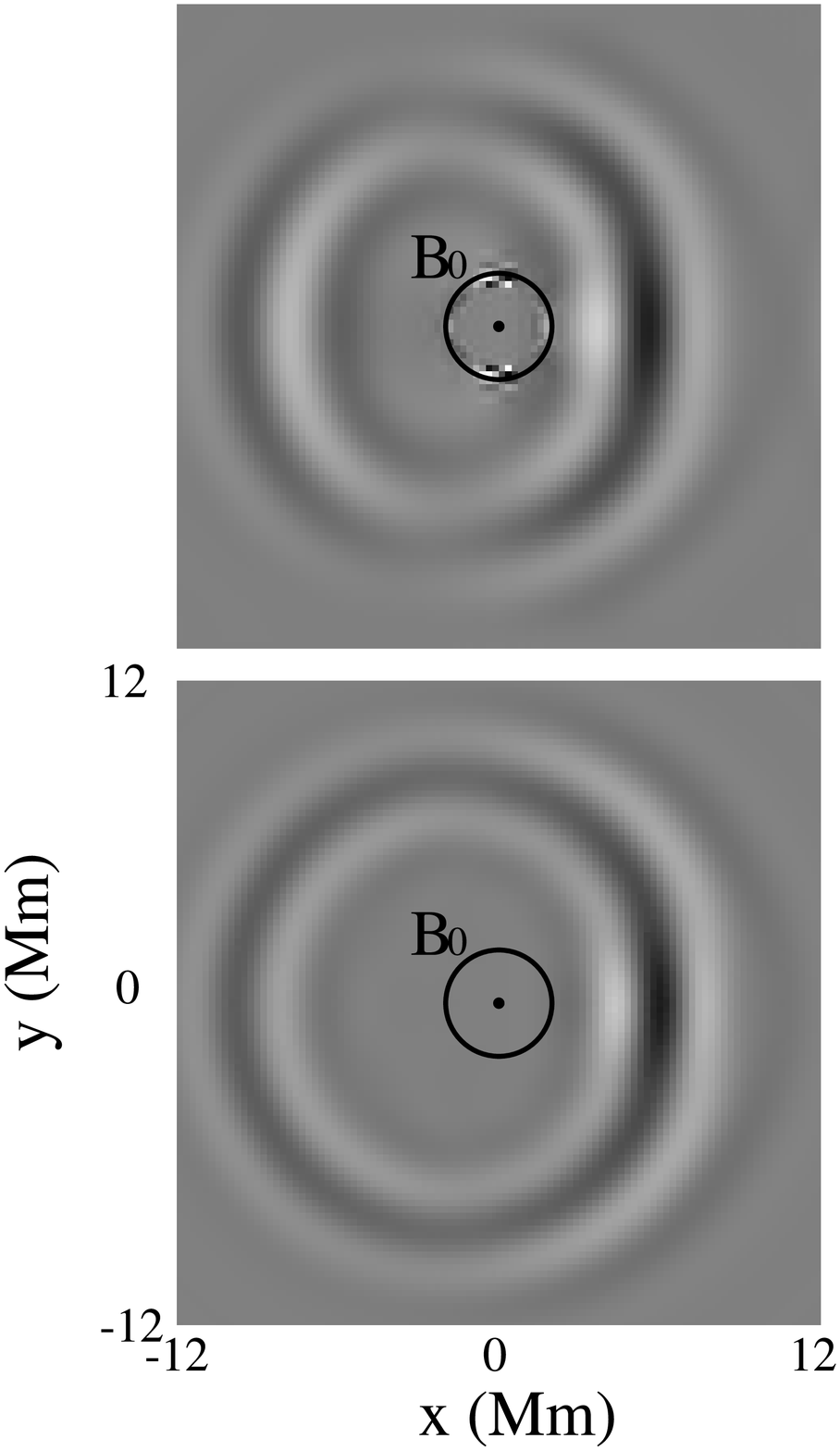}}}
\caption{Scattered pressure perturbations (snapshot) from a strong, 3~kG, vertical flux tube with axis $x=y=0$. The bottom panel shows the exact result, the top panel the numerical. The calculation was performed at
a lower resolution of 100 by 100 grid points with only 3 points across the tube boundary. This shows the type of 
artifacts which can be expected from using a too low resolution.}
\label{fig:h3kG}
\end{figure}

\begin{figure}
\resizebox{0.95\hsize}{!}{\rotatebox{-90}
             {\includegraphics{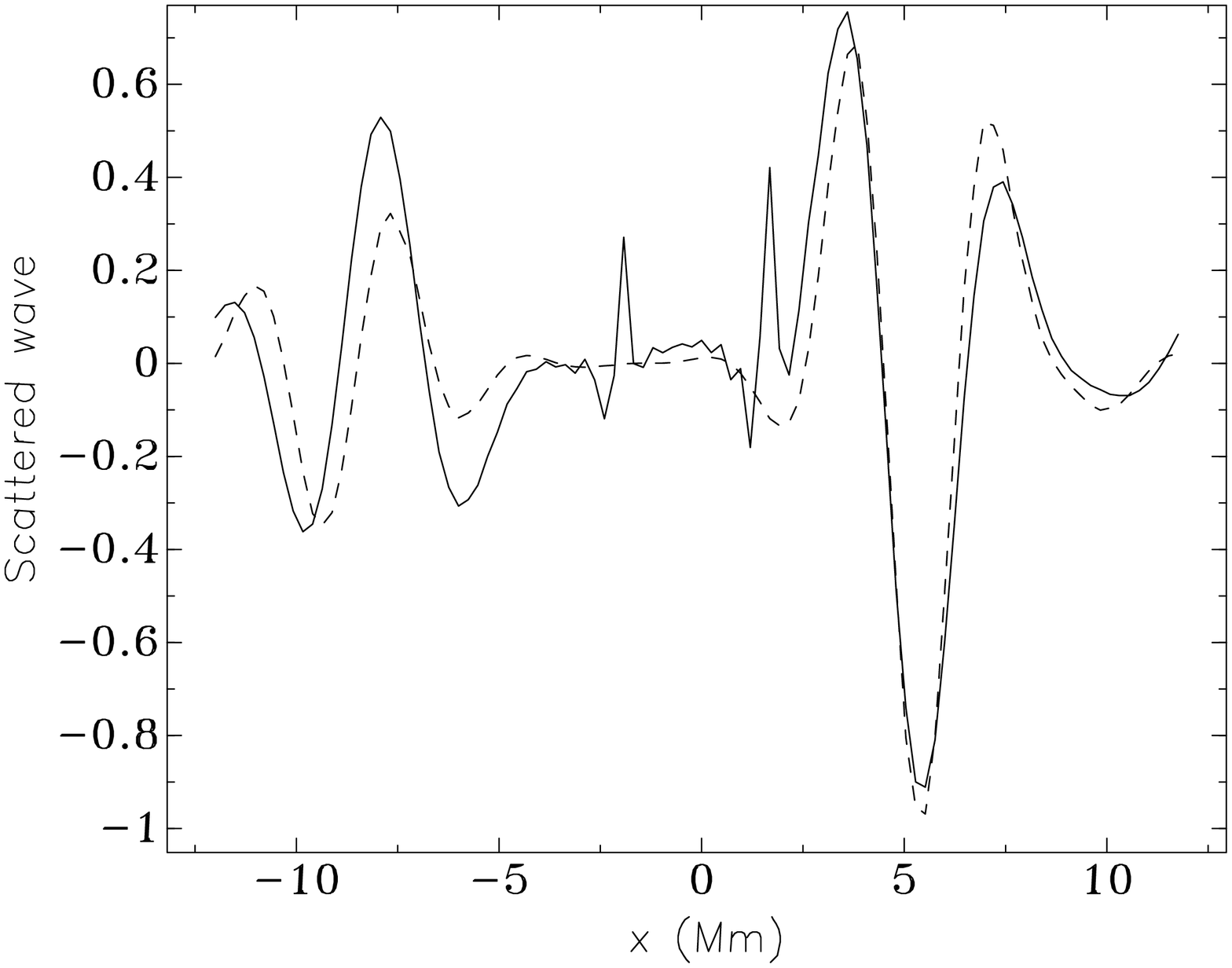}}}
\caption{A cut at $y=0$ through the scattered pressure perturbations shown in Figure~\ref{fig:h3kG} (vertical 3~kG flux tube, low resolution). The dashed line shows the exact solution and the solid line the numerical. The units are such that the incoming pressure perturbations have an amplitude of 1.}
\label{fig:h3kGslice}
\end{figure}

It is also worthwhile to show the results at a lower resolution (here 100
pixels in each direction) with only three points resolving the
boundary layer ($r_1-r_0$). In this case we have chosen a strong 3~kG
magnetic field strength for which
the density in the flux-tube has reduced by between 98\% and 99\%. 
The results are shown in Figures~\ref{fig:h3kG} 
and~\ref{fig:h3kGslice}.
Magnetic fields of this strength, and this level of evacuation, will 
presumably be important in calculations of wave propagation through sunspots. 
Given the resolution, the agreement between the numerical and analytic
results are reasonable, however artifacts near the now poorly resolved 
boundary layer are clearly visible. 

\subsection{Background velocity field}
\begin{figure}
\resizebox{0.95\hsize}{!}{\rotatebox{0}
             {\includegraphics{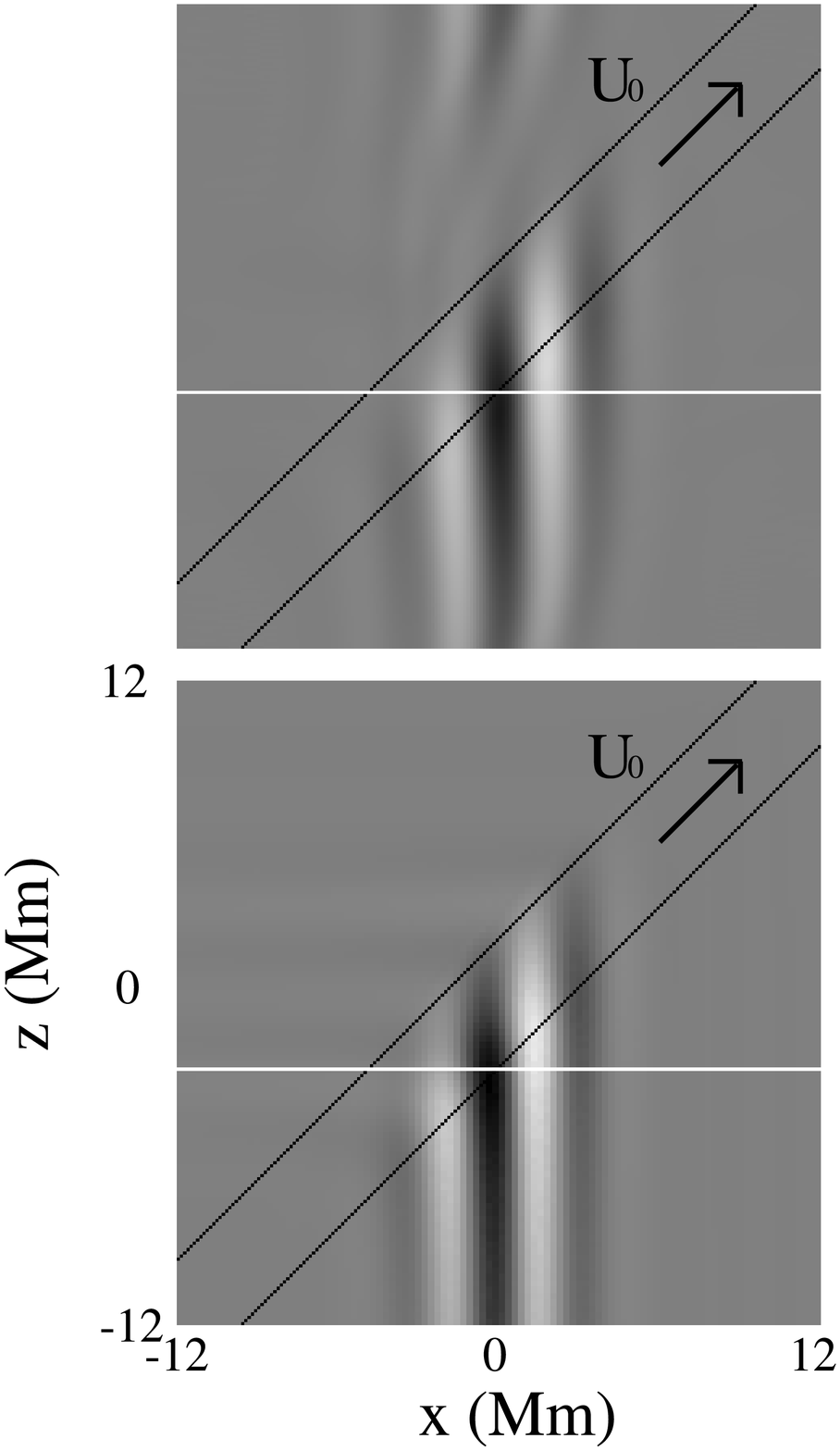}}}
\caption{Scattered pressure perturbations (snapshot) from a 1~km/s flow within a cylinder inclined at $45^\circ$ to the the direction of propagation of the incoming wave. The flow is aligned with the axis of the cylinder, directed upwards and to 
the right. Shown is the
$x$-$z$ plane. The black lines indicate the boundaries of the cylinder.
The bottom panel shows the analytic 
result, the top corresponds to the numerical solution. 
The initial condition for the numerical and analytic treatments are
different, with the analytic treatment considering a wave which has been
in permanent contact with the cylinder. In the top panel, the lower left hand corner is 
least contaminated by the resulting artefacts. Note that because of the periodic nature of the box  scattered waves appear in the top half of the box.
The white line shows the cut used for Figure~\ref{fig:velslice}. }
\label{fig:vel}
\end{figure}

\begin{figure}
\resizebox{0.95\hsize}{!}{\rotatebox{-90}
             {\includegraphics{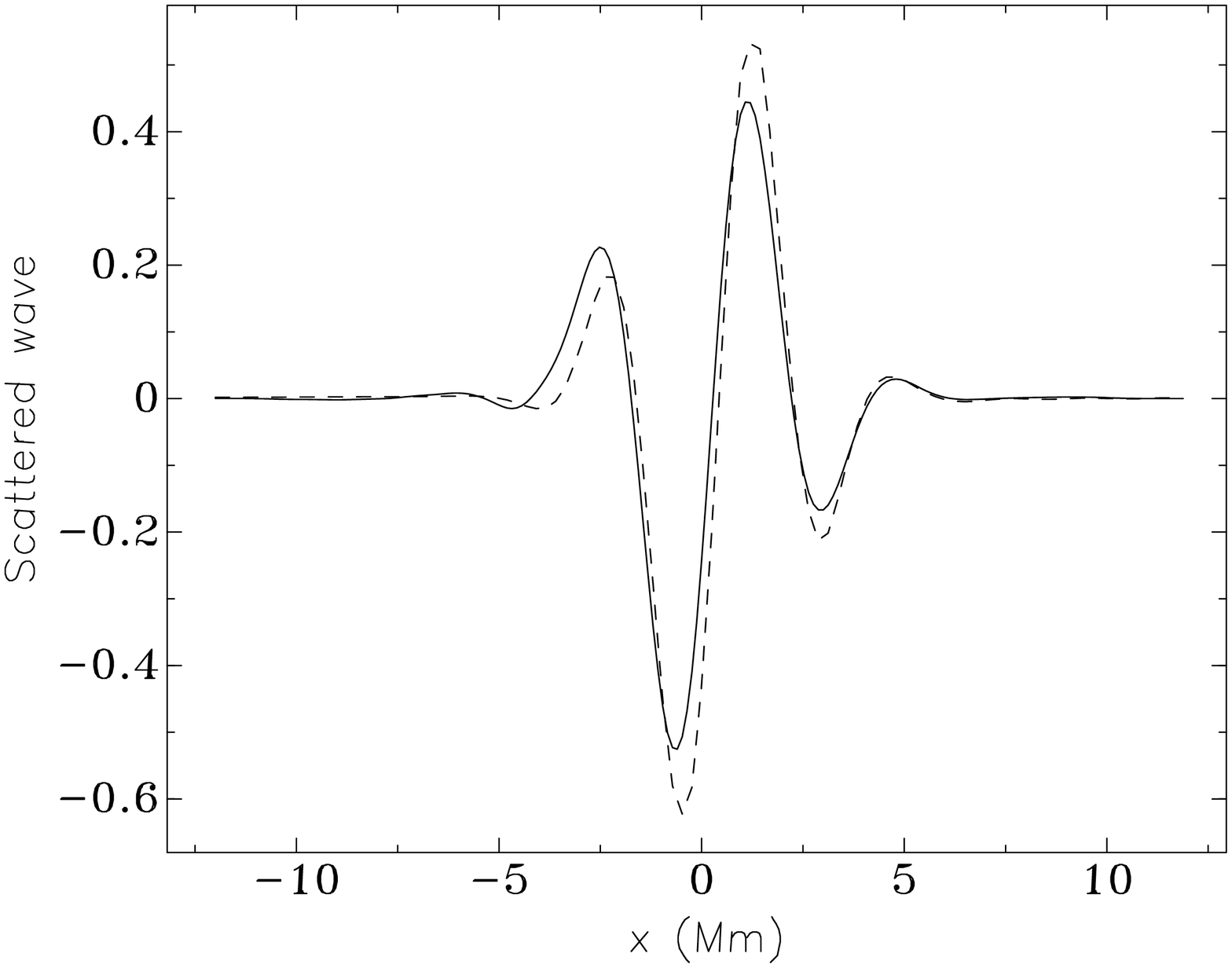}}}
\caption{A cut across  the white lines shown
in Figure~\ref{fig:vel}. The solid curve shows the numerical
solution, the dashed curve the analytic.  }
\label{fig:velslice}
\end{figure}
The above test involves several terms related to background inhomogeneities
in the magnetic field, the density and the pressure fields. An obvious 
remaining class of perturbations involve the velocity field, and we have
tested the code here by comparing against an analytic solution for uniform
flow in a cylinder embedded in a stationary  uniform media. We assume that 
the density and temperature are the same inside and outside the cylinder, 
so that
the only defining property is the velocity which we assume
is directed along the axis of the cylinder. 
The analytic solution was obtained in a similar manner to that in 
Gizon et al. (2006). 

An important property of the solution is that the scattered wave field 
vanishes for a plane parallel wave perpendicular to the cylinder so that all
non-trivial solutions are three-dimensional. We have therefore
chosen to compare the analytic and numerical solutions for the case where
the cylinder and direction of wave propagation make a 45$^{\circ}$ angle.
An additional complication is that the analytic solution no longer
corresponds with an undisturbed plane wave encountering an obstacle:
rather part of the wave-front has previously passed through the
cylinder.

For the calculation we have taken the wave propagation is in the 
$x$ direction with the cylinder running from the bottom left to the 
top right in the $x$-$z$ plane. The radius of the cylinder is 2~Mm, the background
sound speed is 10~km/s, and the flow in the cylinder is 1~km/s. 
Again there is a sharp jump between the properties inside the cylinder
and those outside, and in principle we should use different resolutions
to check the convergence of the solution. In practice we have used a 
grid with 200 grid points or modes in each direction, a similar 
resolution to that found suitable for the magnetic/pressure perturbations.

The analytic and numerical solutions were found to be in reasonable agreement, as is shown in Figures.~\ref{fig:vel} and ~\ref{fig:velslice}. Unlike the 
results shown in section 4.1, an exact 
agreement in this case cannot be expected since the numerical and analytic
problems are significantly different. The analytic
solution is a snapshot of a wave which has always been affected by the tube. 
In contrast the numerical solution evolves, with the scattering beginning at
$t=t_0$. What makes these two solutions a useful test is that the two problems
become identical in the limit of  a large box. 
If we kept increasing the size of our 
box the analytic and numerical solutions would presumably
converge. Tests with smaller boxes (cubes of 12~Mm) indeed showed a much worse
match between the analytic and numerical results, so the agreement we are 
achieving with even 24~Mm boxes is reasonable.

\section{Discussion}
We have written and tested a code for studying wave propagation in an 
inhomogeneous magnetised 
solar atmosphere. There is a wide range of possible applications
of the code, including understanding the observational signature of wave 
propagation through  small-scale magnetic features, through granulation and
supergranulation, and through
the complex magnetic and velocity fields associated with sunspots and their
penumbrae. Tests of the SLiM code so far have been successful. 
We have confidence that the code will be a useful tool to understand 
the signatures of wave propagation through various types of solar features.

\end{document}